\documentclass[aps,onecolumn,nofootinbib,prd,eqsecnum,showpacs,showkeys,preprintnumbers]{revtex4-2}
\usepackage{graphicx}
\usepackage{wrapfig}
\usepackage{caption}
\usepackage{subcaption}
\usepackage{amsmath}
\usepackage{amsfonts}
\usepackage{amssymb}
\usepackage{xcolor}
\usepackage{bm}
\usepackage{natbib}
\usepackage{epstopdf}
\usepackage{url}
\usepackage{footnote}
\usepackage{textcomp}
\usepackage{ulem}
\usepackage{esint}
\usepackage[unicode=true, pdfusetitle,
 bookmarks=true,bookmarksnumbered=false,bookmarksopen=false,
 breaklinks=false,pdfborder={0 0 1},backref=false,colorlinks=false]{hyperref}
\usepackage{multirow}
\usepackage{pifont}
\usepackage{lipsum}
\usepackage[mathlines]{lineno}
\usepackage[newcommands]{ragged2e}
\begin{document}

\title{Constraints on the reheating phase after Higgs inflation in\\
the hybrid metric-Palatini approach}
\author{Brahim Asfour}
\email{brahim.asfour@ump.ac.ma} 
\author{Aatifa Bargach}
\email{a.bargach@ump.ac.ma}
\author{Yahya Ladghami}
\email{yahyaladghami@gmail.com}
\author{Ahmed Errahmani}
\email{ahmederrahmani1@yahoo.fr}
\author{Taoufik Ouali}
\email{t.ouali@ump.ac.ma}
\affiliation{Laboratory of Physics of Matter and Radiation, \\
University of Mohammed first, BP 717, Oujda, Morocco\\
Astrophysical and Cosmological Center, Faculty of Sciences, BP 717, Oujda, Morocco}
\date{\today }

\begin{abstract}
In this paper, we study the post-inflationary era called reheating stage. For this purpose, we consider a model in which the inflaton is non-minimally coupled to the curvature within the hybrid metric-Palatini approach. Furthermore, to investigate the consistency of our results  with the observational data, we relate reheating parameters to those of inflation model. By taking into consideration the Higgs potential $V(\phi)=\lambda/4 \phi^4$; we derive the necessary quantities needed to obtain the reheating duration and the reheating temperature. Moreover, we plot reheating e-folds and temperature as a function of the spectral index, respectively. We consider three cases depending on the coupling constant $\xi$. In addition, we use some specific values of the effective equation of state $\omega$, which is  presumed to remain relatively constant within the range of $-\frac{1}{3} \leq \omega < \frac{1}{3}$. We find that for $\xi=10^{-4.1}$ our results are in agreement with the recent Planck data as the reheating instant is corresponding to the central value of the spectral index and to a maximum temperature required by the scale of baryogenesis models.
\end{abstract}
\maketitle

\section{Introduction}

\par The early cosmic acceleration provides a good solution to several challenges of the hot big bang cosmology (i.e. the flatness, the horizon and the monopole problems). This successful idea of inflation \cite{1, Albrecht, Starobinsky, Linde2, Sato, Linde1,Linde:1990flp} opened new windows in modern cosmology and several inflationary models have been proposed with different classifications, such as single field models (see \cite{Martin:2013tda, Bouabdallaoui:2022wyp, Bargach:2019pst} for more details). 
\par Furthermore, the phenomenological success of scalar field inflation models has the big question of the inflaton’s nature, i.e. which particle physics candidate is able to act as the inflaton field? Untill now the only scalar field that has been detected so far is the Higgs boson \cite{ATLAS:2012yve}. That is why Higgs inflation is one of the most interesting models to study since it makes a connection between the low energy phenomena (Higgs data) and the high-energy physics in the early Universe (inflation). To transcend the major problem with Higgs inflation, as the energy scale of the Higgs field is too small to generate enough e-folds, Non-Minimal Coupling (NMC) between the Higgs field and the Ricci scalar was propesed \cite{Bezrukov:2007ep,Bargach:2020bpf}. Indeed, according to the latest Planck data \cite{Planck:2018jri}, which have severely constrained the values of the scalar spectral index $n_{s}$ and the tensor-to-scalar ratio $r$, inflationary models with the Higgs field NMC to gravity \cite{Bezrukov:2008ej, Rubio:2018ogq} seem to be the most favored.
\par Motivated by the robust reasons behind the Higgs inflationary model, our objective is to study the primary inflationary and reheating eras.  Reheating era \cite{Shtanov:1994ce, Garcia:2020eof, Garcia:2023eol} is the period in which the potential energy of the inflaton is transferred to a thermal bath of matter so that elementary particles start populating the Universe. This makes it the most important application of the quantum theory of particle creation, because during this era the Universe constitutions was created and the Universe gets populated \cite{reheating}. This process which could explain the cosmic origin of matter, has also the responsability of the production of cosmic relics such as photons, neutrinos, and the generation of the observed matter-antimatter asymmetry. The first elementary theory of reheating was developed in \cite{Dolgov:1982th, L.F. Abbott}, and it was also discussed in \cite{Starobinsky2} for the $R^{2}$ inflation model. A reheating phase in the early Universe for the Higgs boson was also considered in \cite{Ahmed:2022tfm} assuming a cubic interaction between the inflaton field and the Standard Model (SM) Higgs boson. As a consequence the reheating is prolonged and the maximal temperature of the SM thermal bath is reduced. Constraints on inflaton Higgs field couplings where studied previously in \cite{Yang:2023rnh} to minimize the probability that the Higgs field entering the unstable regime during reheating.\\
\par Moreover, up to now, restrictions on the reheating energy scale are relatively limited. Clearly, it must be lower than the energy scale of inflation, suggesting that $T_{reh} < 10^{16}$ GeV. Furthermore, the process of reheating must occur prior to Big Bang Nucleosynthesis (BBN), suggesting that $T_{reh} \geq 10$ MeV, for consistency with the standard cosmological model.
\par  Throughout the reheating phase, the cosmic fluid is described by employing an equation of state parameter, denoted as $\omega$. Although the considerable uncertainty surrounding the comprehension of reheating physics, a plausible finite range of values for $\omega$ was suggested. Indeed, besides the canonical reheating scenario which sets $\omega$ equal to 0,   this equation of state should exceed -1/3, as the expansion of the reheating era is not accelerated. Furthermore,  the numerical analyses conducted on the thermalization phase of reheating suggest that its variation is in the range $0 \leq \omega \leq 0.25$  \cite{Podolsky2006}. In this paper, $\omega$ takes specific values in the range $[-1/3,0.25]$ in consistency with the aformentioned fact.
\par In this work, our aim is to study in detail consequences of Higgs dynamics due to NMC Higgs during the reheating phase. In particular, we will explore the implications of the NMC Higgs in the hybrid metric-Palatini model. As it is known, for Einstein gravity we have two different approachs to formulate gravity. The first approach is the metric formalism \cite{Bezrukov:2007ep, Rubio:2018ogq}, where the metric and its first derivatives are the independent variables by imposing the metricity and torsion free conditions from the begining. The second approach is the Palatini formalism \cite{Rasanen:2017ivk, Enckell:2018kkc, Tenkanen:2020dge} in which the metric and the connection serve as independent variables.. The interesting fact here is that predictions differ for different approaches, even if they give the same dynamics for the Einstein gravity. However, once there is a coupling between matter and gravity we obtain different results \cite{Jinno:2019und}. See also \cite{Bauer:2010jg, Rasanen:2018ihz, Rasanen:2017ivk} for predictions for the Higgs inflation parameters in the metric and Palatini formulations of the theory. We notice that the most significant difference between the two formulations is the tensor-to-scalar ratio r, which is predicted to be much smaller in the Palatini formulation \cite{Karam:2020rpa}. Since each formalism suffers from some shortcomings, such as the late-time acceleration of the Universe (dark energy problem) and the gravity quantization challenges, a novel approach by combining elements from both of them has been developed recently and is dubbed as hybrid metric-Palatini formalism \cite{Harko:2011nh, Capozziello}. Over recent years this approach has been extensively researched and refined, see for example \cite{shahid} for Warm inflation in Hybrid metric-Palatini gravity. 
\par The study of Higgs inflation in hybrid metric-Palatini approach has already been conducted in  \cite{Asfour:2022qap} and \cite{He:2022xef}, where the framework considered in the last paper was completely based on a NMC between the Higgs field and both the metric and the Palatini Ricci scalar curvature while in our analysis in \cite{Asfour:2022qap} NMC was exclusively between the Higgs field and the Palatini scalar curvature. The present paper
is an extension of the successful results found in \cite{Asfour:2022qap}, here we attempt to discuss the reheating stage motivated by the increasing interest in Higgs cosmology. So, in this paper, we are interested in reheating stage after inflation of the Universe, where the dynamic is driven by a Higgs field in the context of an hybrid metric-Palatini model. Hence, to check and constrain the reheating era, we adopt the method used in \cite{dai}, where the authors have established relation between inflationary and reheating parameters. After that, we compare our results to the recent observational data provided by Planck experiments \cite{Planck:2018jri}.
\par The outline of the paper is as follows: In section \ref{SecII}, we setup the basic equations of the hybrid metric-Palatini model. In section \ref{secIII}, we compute the duration of reheating besides the reheating temperature. Then, we set constraints to reheating parameters considering Higgs inflation model in section \ref{secIV}. Finally, we conclude and summarize our results in section \ref{SecVI}.

\section{The model} \label{SecII}
In this work, we consider a hybrid metric-Palatini model, in which the inflaton is NMC to the Palatini curvature. The action is described by the following expression \cite{ Asfour:2022qap}  
\begin{equation}
 S=\int d^4x \sqrt{-g}\left[ \frac{1}{2 \kappa^2}R + \frac{1}{2} \xi \phi^2 \hat{R}-\frac{1}{2} \partial_{\mu} \phi \partial^{\mu} \phi - V(\phi)\right], \label{eq2.1}
\end{equation}
where $\kappa^2=M_{pl}^{-2}=8\pi G$ is the reduced Planck mass, $R$ is the Einstein-Hilbert curvature, $\xi$ is the coupling constant, $V(\phi)$ is the scalar field potential, and $\hat{R}$ is the Palatini curvature term. $\hat{R}$ depends on the metric tensor $g_{\mu \nu}$  and on the connection $\Gamma_{\beta \gamma}^{\alpha}$ which are both considered as an independent variables.\\
\par In the Flat Friedmann-Robertson-Walker (FRW) background, we consider the following metric 
\begin{eqnarray}
ds^2=-dt^2+a^2(t)(dx^2+dy^2+dz^2),
\end{eqnarray}
where $a(t)$ is the scale factor and t is the cosmic time. Einstein equations are obtained by varying the action given in Eq.\eqref{eq2.1} with respect to the metric tensor $g_{\mu \nu}$ as follows \cite{ Asfour:2022qap}   
\begin{equation}
    F(\phi)G_{\mu \nu}=\kappa^2T_{\mu \nu}, \label{eq.2.3}
\end{equation}
with
\begin{eqnarray}
 G_{\mu \nu}&=&R_{\mu \nu}-\frac{1}{2}g_{\mu \nu}R,
 \end{eqnarray}
 and
 \begin{eqnarray}
 T_{\mu \nu}&=&A\nabla_{\mu}\phi\nabla_{\nu}\phi-B g_{\mu \nu}(\nabla \phi)^2-g_{\mu \nu}V(\phi)+C\phi\left[ \nabla_{\mu}\nabla_{\nu}-g_{\mu \nu}\square \right] \phi,
\end{eqnarray}

are the Einstein tensor and the energy-momentum tensor, respectively, where the constants A, B and C are given by $A=(1+2\xi-4\xi \sigma)$, $B=\left( \frac{1}{2}+2\xi-\xi \sigma\right)$ and 
$C=2\xi(1+\sigma)$. The parameter $\sigma=0,1$ corresponds to the metric and Palatini cases, respectively. $F(\phi)$ is a function of the scalar field encoding the non-minimal coupling to the Palatini curvature as
\begin{equation}
F(\phi)=1+\xi\kappa^2 \phi^2.
\end{equation}
Considering the $00$-component of Eq.$\eqref{eq2.3}$, we get the Friedmann equation \cite{ Asfour:2022qap}
\begin{equation}
H^2=\frac{\kappa^2}{3F(\phi)}\left[ \left( \frac{1}{2}-3\xi\sigma\right) {\dot{\phi}}^2+V(\phi)-6H\xi(1+\sigma) \phi \dot{\phi}\right], \label{eq2.3}
\end{equation}
where $H=\dot{a}/a$ represents the Hubble parameter, and the dot indicates the derivative with respect to  the cosmic time t. The modified Klein-Gordon equation satisfied by the inflaton field is written as
\begin{equation}
 \square \phi  + \xi \hat{R}\phi - V_{\phi}=0,
\end{equation}
where $ V_{\phi}=dV/d\phi$ is the derivative of the potential with respect to the scalar field $\phi$.\\

By considering the slow roll approximation, Eq.\eqref{eq2.3} can be reduced to \cite{ Asfour:2022qap}
\begin{equation}
H^2	\simeq\frac{\kappa^2 V(\phi)}{3F(\phi)}, \label{eq2.5}
\end{equation}
and the slow roll parameters are given by 
\begin{eqnarray}
 &\epsilon&=\frac{1}{2\kappa^2}\left( \frac{V_{\phi}}{V}\right)^2 Q, \label{eq2.7} \\
  &\eta&=\frac{ V _{\phi\phi}}{3H^2},\\
 &\zeta&=6\xi\sigma,\\
 &\chi&=\kappa^2_{eff}\left(  (1+2\xi-4\xi\sigma)\dot{\phi}-2\xi(1+\sigma)H\phi\right)
 \frac{(1+4\xi -2\xi\sigma)\dot{\phi}+10\xi(1+\sigma) H\phi}{2 F(\phi) H^2},
\end{eqnarray}
where the correction term $Q$ is defined as 
 \begin{equation}
 Q=\frac{F}{\alpha}\left( 1-\frac{4\xi\kappa^2\phi}{F(\phi)}\frac{V}{V_{\phi}}\right) \left( 1-\frac{2\xi\kappa^2\phi}{F(\phi)}\frac{V}{V_{\phi}}\right),
 \end{equation}
 with $\alpha=1-6\xi\sigma$. The standard slow roll parameters \cite{Liddle:1994dx} are recovered when $\alpha=1$. Another important parameter that characterize the model is the number of e-folds defined as
\begin{equation}
 N=\int_{t_k}^{t_{end}}Hdt=\int_{\phi_k}^{\phi_{end}}\frac{H}{\dot{\phi}}d\phi, \label{eq2.12}
\end{equation}
where the subscript "$k$" and "end" represent the crossing horizon and the end of inflation, respectively.\\
The power spectrum of the curvature perturbations and the tensor perturbations amplitude are given, respectively, by the following expressions \cite{ Asfour:2022qap},
\begin{eqnarray}
   A^2_s&=&\frac{4}{25}P_R=\frac{\kappa^6 V^3}{75\pi^2V^2_{,\phi}}Q_{2},
\end{eqnarray}
and
\begin{eqnarray}
A^2_T=\frac{2\kappa^2}{25}(\frac{H}{2\pi})^2=\frac{4\kappa^4 V}{600\pi^2}Q_{3},
\end{eqnarray}
where the correction terms $Q_{2}$ and $Q_{3}$ are
\begin{eqnarray}
    Q_{2}=\frac{(1-6\xi\omega)^2}{F(\phi)\left[ 1+\frac{C\kappa^2}{2F(\phi)H}\dot{\phi}\phi\right]^2}\frac{V^2_\phi}{(2F_{,\phi}V-F(\phi) V_{,\phi})^2},
\end{eqnarray}
and
\begin{equation}
    Q_{3}=\frac{1}{F(\phi)},
\end{equation}
respectively.
The spectral index of the power spectrum and the tensor-to-scalar ratio can be formulated in terms of slow roll parameters as follows \cite{ Asfour:2022qap}
\begin{equation}
  n_s=1-2\epsilon+\frac{2}{\alpha}\left(\eta-\frac{\zeta}{3}-2\chi\right), \label{eq2.13}
\end{equation}
and
\begin{eqnarray}
\nonumber r&=&\dfrac{A^2_T}{A^2_S}\\
&=&\frac{1 }{2\kappa^2 \alpha^2}\frac{V^2_\phi}{ V^2} \left[ 1+\frac{C\kappa^2}{2F(\phi)H}\dot{\phi}\phi\right]^2, \label{eq5.6}
\end{eqnarray}
respectively.
\section{Reheating} \label{secIII}
It is important to note that it is difficult to constrain the reheating phase by observational data. Then, as we have previously mentioned, we need to adopt an approach to find a relation between the reheating and the inflationary parameters. In the aim of extracting information about reheating, i.e. the duration and the temperature of reheating, we consider a comoving wave number mode $k$, which crosses the horizon during inflation at $a=a_{k}$. Throught the following equation, we can relate the comoving Hubble scale at the horizon crossing, $a_{k}H_{k}=k$, to that of the present time as  \cite{dai}
\begin{equation}
\frac{k}{a_{0}H_{0}}=\frac{a_k}{a_{end}}\frac{a_{end}}{a_{re}}\frac{a_{re}}{a_0}\frac{H_k}{H_{0}}, \label{eq3.1}
\end{equation}
where $a_{0}$, $a_k$, $a_{re}$, and $a_{end}$ denote respectively, the scale factor at the present time, the horizon crossing time, the end of reheating, and the end of inflation, while $H_{0}$ represents the present Hubble constant.        
Using $N_k=\ln(a_{end}/a_k)$, and $N_{re}=\ln(a_{re}/a_{end})$, we obtain the following relation
\begin{equation}
\ln\frac{k}{a_{0}H_{0}}=-N_k-N_{re}+ln\frac{a_{re}}{a_0}+ln\frac{H_k}{H_{0}}, \label{eq3.2}
\end{equation}
where $N_k$ is the number of e-folds between the time when a mode exits the horizon and the end of inflation, while $N_{re}$ is the number of e-folds between the end of inflation and the time at the end of reheating. Assuming no entropy production after the completion of reheating, one can write \cite{creminelli2014phi,koh2018constraints}
\begin{eqnarray}
\frac{a_{re}}{a_0}=\frac{T_0}{T_{re}}\left(\frac{43}{11g_{*}(T_{re})}\right)^{1/3}, \label{eq3.3}
\end{eqnarray} \linebreak
where $T_0$ is the current temperature of the Universe, $T_{re}$ is the thermal equilibrium temperature of reheating, and $g_{*}(T_{re})$ is the number of relavistic degrees of freedom at the end of reheating. 
The energy density at the end of reheating, $\rho_{re}$, can be written in terms of the reheating temperature as follows \cite{koh2018constraints}
\begin{equation}
\rho_{re}=\frac{\pi^2}{30}g_{*}(T_{re})T_{re}^4. \label{eq3.4}
\end{equation}
Using $\rho\propto a^{-3(1+\omega)}$, one can set a relation between the energy density at the end of reheating and the energy density at the end of inflation (the beginning of reheating), $\rho_{end}$, as    
\begin{equation}
\frac{\rho_{re}}{\rho_{end}}= \left(\frac{a_{re}}{a_{end}}\right)^{-3(1+\omega)},
\end{equation}   
where the effective equation of state parameter, $\omega$, is assumed to be constant during reheating era. By including the number of e-folds $N_{re}$, we obtain \cite{munoz2015equation}
\begin{equation}
\rho_{re}=\rho_{end}e^{-3(1+\omega)N_{re}}. \label{eq3.6}
\end{equation}
Furthermore, $\rho_{end}$ is related to the potential at the end of inflation $V_{end}$ as follows :
\begin{equation}
\rho_{end}=\lambda_{end}V_{end}, \label{eq3.7}
\end{equation}
where $\lambda_{end}$ is the effective ratio of the kinetic energy to the potential energy at the end of inflation. Using Eqs. \eqref{eq3.2}-\eqref{eq3.4}, and \eqref{eq3.7}, we find the duration of reheating \cite{koh2018constraints}
\begin{equation}
N_{re}= \frac{4}{1-3\omega}\left[-\ln\left(\frac{k}{a_0 T_0}\right)-\frac{1}{3}\ln\left(\frac{11g_{*s}}{43}\right)\\
-\frac{1}{4}\ln\left(\frac{30\lambda_{end}}{\pi^2g_{*}}\right)-\frac{1}{4}\ln\left(\frac{V_{end}}{H_{k}^4}\right)-N_k \right], \label{eq3.8}
\end{equation}
where we have assumed that the EoS is not equal to 1/3, as $N_{re}$ is not defined for this value of $\omega$. Considering the following numerical values $a_0=1$, $T_0=2.725$ K, $g_{*}=106.75$, $\kappa^{-1}=2.435\times10^{18}$, $k=0.05$ Mpc$^{-1}$ \cite{Planck:2018jri}, Eq.\eqref{eq3.8} becomes
\begin{widetext}
\begin{equation}
N_{re}= \frac{4}{1-3\omega}\left[60.0085-\frac{1}{4}\ln\left(\frac{30\lambda_{end}}{100 \pi^2}\right)-\frac{1}{4}\ln\left(\frac{V_{end}}{H_{k}^4}\right)-N_k \right]. \label{eq3.9}
\end{equation}
\end{widetext}

From Eqs. \eqref{eq3.4}, \eqref{eq3.6} and \eqref{eq3.7}, we can derive the reheating temperature as

\begin{equation}
T_{re}=\left(\frac{30 \lambda_{end} V_{end}}{\pi^2 g_{*}}\right)^{1/4}e^{-\frac{3}{4}(1+\omega)N_{re}}. \label{eq3.10}
\end{equation}
It is evident from this expression that the reheating temperature reaches its maximum when $N_{re}=0$, which means that the reheating happens instantaneously after the end of inflation. In the other hand, our main goal in this work is to constrain the essential reheating parameters $N_{re}$ and $T_{re}$ obtained in Eqs.\eqref{eq3.9} and \eqref{eq3.10}, respectively. To do so, we need to calculate the inflationary quantities $V_{end}$, $N_k$, and $H_{k}$, which  depend on the model under consideration.
\section{Reheating constraints in Higgs inflation} \label{secIV}
\subsection{Higgs inflation}
\begin{figure}[h!]
\centering
\includegraphics[scale=0.5]{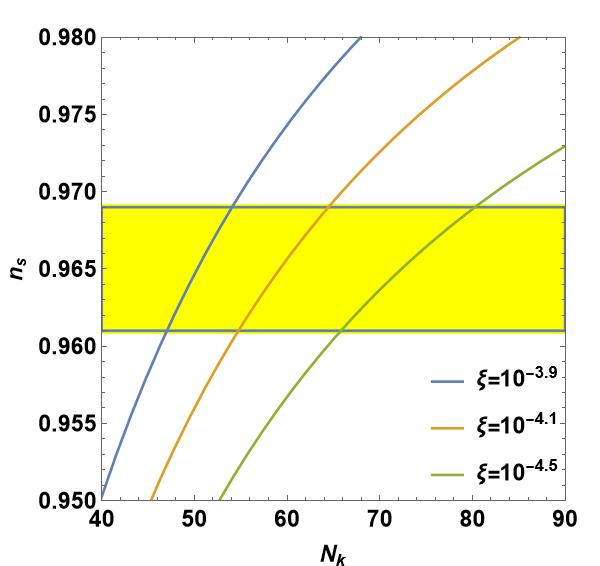}  
\caption{Plot of the spectral index as a function of the number of e-folds for $\xi=10^{-3.9}$, $10^{-4.1}$ and $10^{-4.5}$.}
\label{Fig:1}
\end{figure}

To illustrate our purpose, we consider the Higgs inflation model, where the potential is given by \cite{ Asfour:2022qap,Bargach:2020bpf}
\begin{equation}
V(\phi)=\frac{1}{4}\lambda \phi^4. \label{eq4.1}
\end{equation}
In this study, as we work in the large field regime \cite{Asfour:2022qap}, an extremely small value of the self-coupling constant $\lambda=10^{-9}$ is assumed \cite{Granda:2019wip}. The number of e-folds and the spectral index, given by Eqs \eqref{eq2.12} and \eqref{eq2.13}, take the form
 \begin{eqnarray}
N_k=\frac{\alpha\kappa^2}{8}\left[ \phi_k^2-\phi_{end}^2\right], \label{eq4.2}
\end{eqnarray}
and
\begin{eqnarray}
 n_s=&1&- \frac{16}{\alpha \kappa^2 \phi^2} (1 - \frac{\xi \kappa^2  \phi^2}{2 F(\phi)})\\
 &+&\frac{2}{\alpha } [\frac{12 F(\phi)}{\kappa^2  \phi^2} - 2 \xi \sigma - \kappa_{eff} ((1 + 2 \xi - 4 \xi \sigma) \dot{\phi} - 2 \xi (1 + \sigma) H \phi)\frac{ (1+4\xi -2\xi \sigma) \dot{\phi} +10 \xi(1+ \sigma) H \phi}{F(\phi) H}], \label{4.3}
\end{eqnarray}
respectively. Fig. \ref{Fig:1} shows the variation of the spectral index $n_s$ as a function of the e-folding number $N_k$ for $\xi=10^{-3.9}$, $\xi=10^{-4.1}$ and $\xi=10^{-4.5}$. The yellow horizontal region indicates the bound imposed on the spectral index by the observational data i.e. $n_s= 0.9649 \pm 0.0042$ \cite{Planck:2018jri}. It can be seen that the results are consistent with the observational data for an appropriate range of $N_k$.

At the end of inflation, using Eq. \eqref{eq2.7}, we obtain
\begin{equation}
\phi_{end}=\left[\frac{\alpha-4\xi}{2\xi\kappa^2\alpha}\left(\sqrt{1+\frac{32\xi\alpha}{\alpha-4\xi}}-1\right)\right]^{\frac{1}{2}}. \label{eq4.4}
\end{equation}
One can compute $V_{end}$ from Eqs. \eqref{eq4.1} and \eqref{eq4.4} as 
\begin{equation}
V_{end}=\frac{\lambda}{4}\left[\frac{\alpha-4\xi}{2\xi\kappa^2\alpha}\left(\sqrt{1+\frac{32\xi\alpha}{\alpha-4\xi}}-1\right)\right]^2,
\end{equation}
and $\lambda_{end}$, from Eqs. \eqref{eq2.3} and \eqref{eq3.7}, as
\begin{equation}
\lambda_{end}=\left[\left(\frac{1}{2}-3\xi\sigma\right)\frac{\dot{\phi}^2}{V(\phi)}+1-\frac{6\xi(1+\sigma)H\phi\dot{\phi}}{V(\phi)} \right]_{\phi=\phi_{end}}.
\end{equation}
From Eq. \eqref{eq4.2}, we find the inflaton field at the horizon crossing $\phi_k$ in terms of $N_k$ as
\begin{equation}
(\kappa \phi_k)^2=\frac{8}{\alpha}\left[N_k+\frac{\alpha-4\xi}{16\xi}\left(\sqrt{1+\frac{32\xi\alpha}{\alpha-4\xi}}-1\right)\right].
\end{equation}
By incorporating this expression into the reduced Friedmann equation Eq. \eqref{eq2.5}, we obtain the Hubble parameter at the time of horizon crossing as

\begin{equation}
H_k^2=\frac{16\lambda}{3\kappa^2 \alpha^2F(\phi)}\left[N_k+\frac{\alpha-4\xi}{16\xi}\left(\sqrt{1+\frac{32\xi\alpha}{\alpha-4\xi}}-1\right)\right]^2.
\end{equation}
\subsection{Reheating constraints}
\subsubsection{Reheating duration}
\begin{figure}[h!]
\centering
\begin{subfigure}[b]{0.49\textwidth}
\includegraphics[height=2.6in]{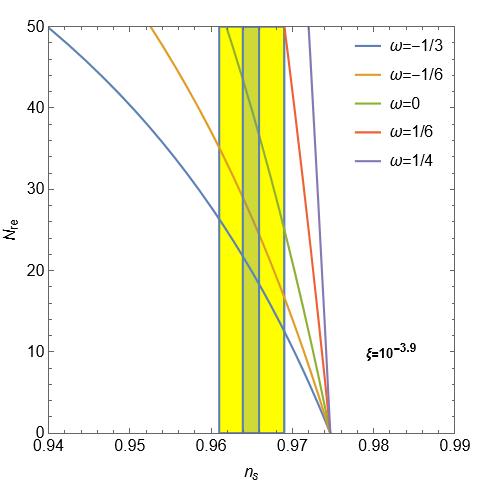}
\caption*{-2a-}
\end{subfigure}
\begin{subfigure}[b]{0.49\textwidth}
\includegraphics[height=2.6in]{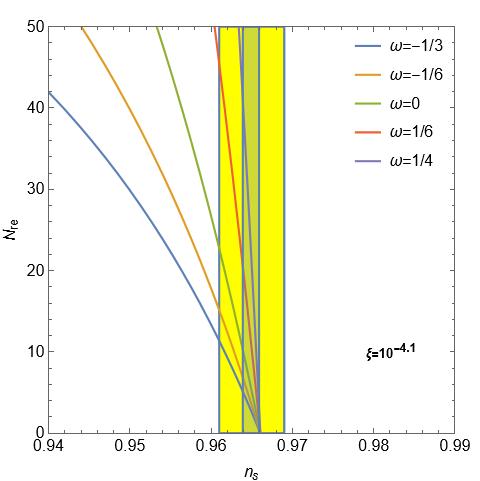} 
\caption*{-2b-}
\end{subfigure}
\begin{subfigure}[b]{0.49\textwidth}
\includegraphics[height=2.6in]{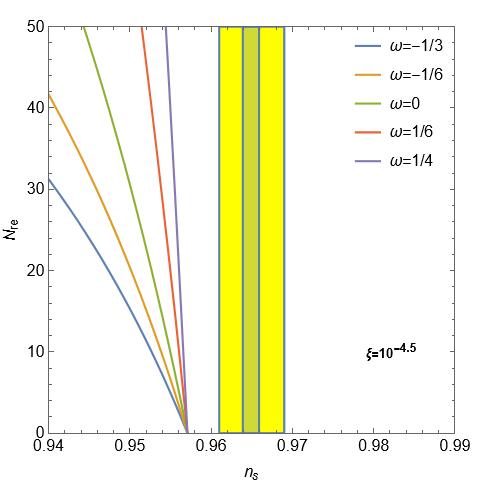}  
\caption*{-2c-}
\end{subfigure}
\caption{Variation of the reheating duration as a function of the spectral index for different values of $\omega$, considering three values of the coupling constant $\xi=10^{-3.9}$, $10^{-4.1}$ and $10^{-4.5}$.}
\label{Fig:2}
\end{figure}

Using the results of the previous subsection, we plot in Fig. \ref{Fig:2} the variation of the reheating duration as expressed in Eq. \eqref{eq3.9} as a function of the specral index given by Eq. \eqref{4.3}, for different values of the effective equation of state i.e. $\omega=-1/3$ (the blue curve), $\omega=-1/6$ (the orange curve), $\omega=0$ (the green curve), $\omega=1/6$ (the red curve) and $\omega=1/4$ (the purple curve). The Figure is plotted by taking into consideration three values of the coupling constant $\xi=10^{-3.9}$ (on the left panel), $\xi=10^{-4.1}$ (on the  right panel) and $\xi=10^{-4.5}$ (on the  lower panel). The vertical yellow region indicates the bound imposed by Planck on $n_s$, i.e. $n_s= 0.9649 \pm 0.0042$  \cite{Planck:2018jri}, and the dark green region represents a precision of $10^{-3}$ from future observations \cite{Amendola:2016saw}. We observe that all curves intersect at one point where $N_{re}\rightarrow0$, corresponding to the reheating instant. In the case where $\xi=10^{-4.1}$, we obtain the most compatible results with the Planck data, as all curves shift towards the central value of the spectral index $n_s=0.9649$. For $\xi=10^{-3.9}$, the results can be consistent with observations but require a more exotic mechanism of reheating. It is difficult to reconcile with the bounds imposed on $n_s$ in the case where $\xi=10^{-4.5}$ for all values of $\omega \leq \frac{1}{4}$.
\captionsetup[figure]{justification=Justified}

\subsubsection{Reheating temperature}
\begin{figure}[h!]
\centering
\begin{subfigure}[b]{0.49\textwidth}
\includegraphics[height=2.6in]{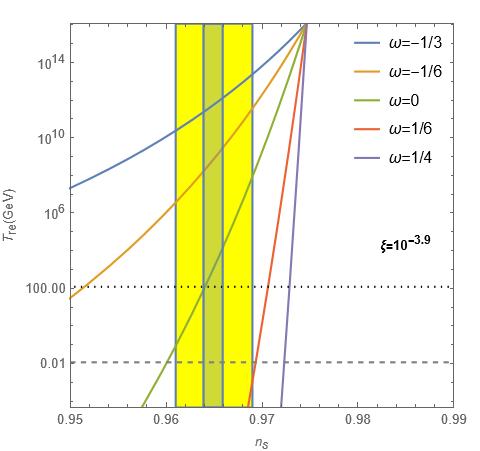} 
\caption*{-3a-}
\end{subfigure}
\begin{subfigure}[b]{0.49\textwidth}
\includegraphics[height=2.6in]{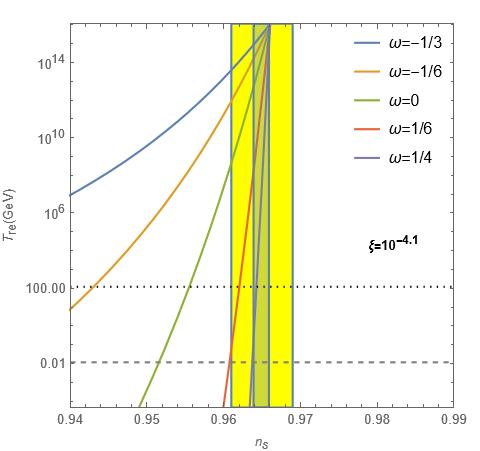}   
\caption*{-3b-}
\end{subfigure} 
\begin{subfigure}[b]{0.49\textwidth}
\includegraphics[height=2.6in]{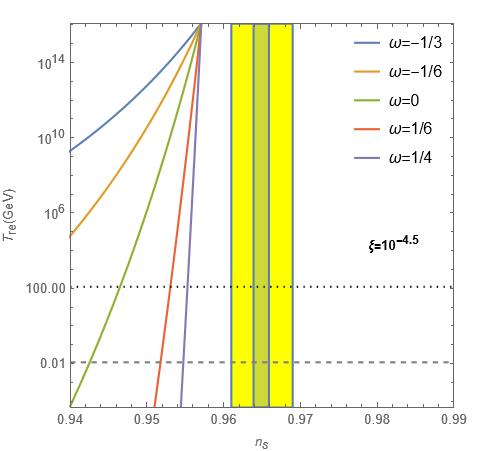} 
\caption*{-3c-}
\end{subfigure} 
\caption{Evolution of the reheating temperature versus the spectral index for different values of $\omega$, considering three values of the coupling constant $\xi=10^{-3.9}$, $\xi=10^{-4.1}$ and $\xi=10^{-4.5}$.}
\label{Fig:3}
\end{figure}

Fig. \ref{Fig:3} shows the variation of reheating temperature $T_{re}$ as a function of the spectral index $n_s$, using the same values of the equation of state $\omega$ and coupling constant $\xi$ considered in Fig. \ref{Fig:2}. The horizontal dashed lines at $T_{EW}=10^2$ GeV (black) and $T_{re}=10$  MeV (gray) indicate the electroweak (EW) scale and the big bang nucleosynthesis scale, respectively. It's observed that all curves converge to a single point where the temperature is very large and maximal i.e. $T_{re}=10^{16}$ GeV as may be required by grand unification scale baryogenesis models. Additionally, it's noteworthy that this point corresponds to instantaneous reheating i.e. $N_{re}\longrightarrow0$. Temperatures that lie above this intersection point are considered unphysical as their results yield a negative value for $N_{re}$.

\captionsetup[figure]{justification=Justified}

\section{Conclusion} \label{SecVI}

In this paper, we have investigated the phase that follows the inflation era, namely the reheating phase. To do so, we have considered the framework of non-minimal coupling in the hybrid metric-Palatini approach, where we have chosen Higgs field as an inflaton.\\
\par To this aim, we have started by reviewing basic equations of the hybrid metric-Palatini model with non-minimal coupling between gravity and the scalar field. We have calculated the reheating duration, as well as the temperature and we have also related these two quantities to the inflationary parameters. This step has allowed us to constraint the reheating era by the observational data imposed on the inflation one \cite{Planck:2018jri}.
\par The aforementioned study was illustrated in the case of the Higgs inflation model. We have computed crucial parameters such as $n_s$, $V_{end}$, $\lambda_{end}$, and $H_k$. Subsequently, we have set constraints on the reheating by plotting the number of e-folds $N_{re}$, Fig.\eqref{Fig:2}, as well as the temperature $T_{re}$, Fig.\eqref{Fig:3}, within this phase as a function of the spectral index $n_s$. To do this, we have chosen specific values of the equation of state $\omega$ and we have considered three values of the coupling constant as shown in all figures. We have found that the temperature at the end of reheating reaches its maximum value i.e. $T_{re}=10^{16}$ GeV for the instantaneous reheating.
\par We conclude that the non-minimal coupling between gravity and scalar field in the Higgs hybrid metric-Palatini model is consistent with the recent observational data \cite{Planck:2018jri}. Notably, for the coupling constant value $\xi=10^{-4.1}$, the reheating instant is corresponding to the central value of the spectral index and to a maximum temperature required by the grand unification scale baryogenesis models.
\par In the upcoming works, to enhance our comprehension and to obtain a thorough understanding, we are going to investigate in the same framework the preheating era, which is the phase that precede the reheating one. We will also study the production of the primordial gravitational waves as well as the primordial black holes during the early Universe in the framework of hybrid metric-Palatini approach.

\end{document}